\documentclass[3p,times]{elsarticle}

\usepackage{ecrc}
\usepackage{url}
\usepackage{amsmath}
\usepackage[T1]{fontenc}
\usepackage{epstopdf}
\usepackage{multirow}
\usepackage{color}
\volume{00}

\firstpage{1}

\journalname{Journal of Informetrics}

\runauth{Mayank Singh et al.}

\jid{procs}

\jnltitlelogo{Journal of Informetrics}

\CopyrightLine{2011}{Published by Elsevier Ltd.}

\usepackage{amssymb}
\usepackage[figuresright]{rotating}

\begin{document}

\begin{frontmatter}

\dochead{}

\title{Is this conference a top-tier? ConfAssist: An assistive conflict resolution framework for conference categorization}

\author{Mayank Singh, Tanmoy Chakraborty, Animesh Mukherjee, Pawan Goyal}

\address{Department of Computer Science and Enginnering\\
	 Indian Institute of Technology, Kharagpur, WB, India\\
	 \{mayank.singh,its\_tanmoy,animeshm,pawang\}@cse.iitkgp.ernet.in}

\begin{abstract}
Classifying publication venues into top-tier or non top-tier is quite
subjective and can be debatable at times. In this paper, we propose
\textit{ConfAssist}, a novel assisting framework for conference
categorization that aims to address the limitations in the existing
systems and portals for venue classification. We start with the
hypothesis that top-tier conferences are much more stable than
other conferences and the inherent dynamics of these groups differs to
a very large extent. We identify various features related to
the stability of conferences that might help us separate a top-tier
conference from the rest of the lot. While there are many 
clear cases where expert agreement can be almost immediately achieved as to whether 
a conference is a top-tier or not, there are equally many cases that can result in a conflict 
even among the experts. \textit{ConfAssist} tries to serve as an aid in such cases by 
increasing the confidence of the experts in their decision.
An analysis of 110 conferences from 22 sub-fields of computer science clearly favors our
hypothesis as the top-tier conferences are found to exhibit much less
fluctuations in the stability related features than the non top-tier ones. We
evaluate our hypothesis using systems based on conference
categorization.  For the evaluation, we conducted human judgment survey with 28 domain experts. The results are impressive
with 85.18\% classification accuracy. We also compare the dynamics of the newly started conferences with the older conferences to identify the initial signals of popularity. The system is applicable to any conference with atleast 5 years of publication history.
\end{abstract}


\end{frontmatter}

\section{Introduction}
Conferences are accepted as the primary means to communicate the
results, ideas and innovation among the computer science research
community. Researchers always prefer to present their work in the best
venues to get their work recognized among the peers of the field. A
very interesting research question arises here, that is, is it
possible to identify certain features that can distinguish top-tier
venues from the rest of the lot? More importantly, if we can identify
some non-trivial features apart from the raw citation counts, that serve
as the distinguishing factors, it can throw some light on a more
fundamental problem, that is, what takes to become a top-tier
conference? 

The scientific community has always been demanding for better
algorithms, metrics and features for scientific venue ranking and
categorization. Different organizations, researchers and forums
provide different
rankings\footnote{http://scholar.google.co.in.}\textsuperscript{,}\footnote{http://academic.research.microsoft.com/.}\textsuperscript{,}\footnote{http://arnetminer.org.}\textsuperscript{,}\footnote{http://www.scimagojr.com/journalrank.php.}\textsuperscript{,}\footnote{http://admin-apps.webofknowledge.com/JCR/JCR.}
and categorization of venues\footnote{http://www.ntu.edu.sg/home/assourav/crank.htm.}\textsuperscript{,}\footnote{http://webdocs.cs.ualberta.ca/~zaiane/htmldocs/ConfRanking.html.}\textsuperscript{,}\footnote{http://perso.crans.org/~genest/conf.html.}\textsuperscript{,}\footnote{http://portal.core.edu.au/conf-ranks/.}\textsuperscript{,}\footnote{http://dsl.serc.iisc.ernet.in/publications/CS\_ConfRank.htm.}. 

The existing systems and portals for venue classification, however,
have several limitations. First, most of the existing systems provide
category-based rankings, but no clear demarcation between these categories. Second, existing
systems that provide category based classification fail to provide the
main intuitions behind such classification. Third, ranking systems
use h-index\footnote{http://scholar.google.co.in.}\textsuperscript{,}\footnote{http://academic.research.microsoft.com/.} and impact factor based metrics~\footnote{http://www.cs.iit.edu/~xli/CS-Conference-Journals-Impact.htm.}\textsuperscript{,}\footnote{http://www.scimagojr.com/journalsearch.php?q=conference.}, which in turn are very debatable \citep{Sekercioglu2008,EMBR:EMBR200974,labbe:hal-00713564,meyer2009viewpoint}. 
Fourth, almost all such systems are domain dependent \citep{doi:10.1108/00220410810844150}.
We address some of these limitations in this work by proposing some quantitative measures that are largely unbiased by raw citation counts.

Each venue aims to maximize its citation counts and come into the
league of top-tiers. Similarly, those who are already in this league,
strive to maintain their standards. There could be several underlying
parameters other than the citation counts, that reflect the standard
of a conference. For instance, can top-tier conferences be
distinguished from the rest of the lot based on the observation that
the top-tier conferences do not tend to undergo a drastic change in some characteristics/parameters over
the years? In this paper, we attempt to develop a high confidence
venue classification system. While the experts might agree on the very clear cases, there might be
cases of conflict and such a system can assist in making a decision (see results in Table \ref{tab:cat-survey-agree}).

In this paper, we demonstrate the development of \textit{ConfAssist} which is
a novel conflict resolution framework that can assist experts to resolve conflicts in deciding
whether a conference is a top-tier or not by expressing how (dis)similar the conference
is to other well accepted top-tier/ non top-tier conferences. The contributions of the work are as follows:
\textit{i) Motivation: }
We start with an experiment that shows while there are many 
clear cases where expert agreement can be almost immediately achieved as to whether 
a conference is a top-tier or not, there are equally many cases that can result in a conflict 
even among the experts. We also demonstrate that high impact factor and low acceptance rate are not always proportional to the popularity of the conferences.
\textit{ii) The hypothesis: }
We present a hypothesis that the top-tier
conferences are much more stable in terms of maintaining the same
level of diversity over the years. To test our hypothesis, we explore
the conference dynamics by analyzing how much various
parameters fluctuate over the years for the top-tier conferences as
compared to the non top-tier ones. For this study, we take
110 conferences from 22 sub-fields of the computer science domain.
\textit{iii) The features: } The conference
parameters that we explore using our hypothesis include various
diversity patterns such as (1) diversity in terms of the computer
science fields referred to by the papers appearing in a conference,
(2) diversity of the keywords used in the conference papers, (3)
diversity in terms of the research fields of the authors publishing in
the conference, (4) proportion of new authors and (5) diversity in the
publication age of authors.  For the accepted papers in the
conference, we also use some features from the co-authorship network
between the authors of the accepted papers. These features include (6) diversity in degree of author node, (7) diversity
in edge strength of author-author link, (8) diversity in average
closeness centrality and (9) diversity in average betweenness
centrality. We use the fluctuations observed in these parameters as
features to categorize conferences into top-tier or non top-tier.
\textit{iv) The evaluation: }We evaluate our hypothesis using systems
based on conference categorization. For this, we conduct a human
judgment survey with 28 domain experts. We achieve as high as 85.18\% classification accuracy. 

The work presented here is an extension of \cite{Singh:2015}. The novel aspects and contributions of this paper with respect to the 
conference version are: a) It is an extended version of the conference poster with a detailed explanation of the proposed system
as well as the features utilized, b) We report a detailed feature as well as parameter analysis in this paper. 
The rest of the paper has been organized as
follows. We discuss the related previous works in Section
\ref{sec:relWork}. Section \ref{sec:motivation} describes analysis of conference-level data for motivational experiments that form the basis for this study.  
Section \ref{sec:dataset} describes the dataset
used for our experiments. Various features utilized for our study have been
described in Section \ref{sec:features}. A detailed analysis of
various features has been presented in Section \ref{sec:analysis},
where we also do a field based comparison. In this section, we further
compare the dynamics of newly starting conferences with older top-tier and
non top-tier conferences to identify initial signals of
popularity. The experiments to evaluate our system under different
settings have been reported in Section
\ref{sec:experiments}. A factor analysis to determine the different latent factors in the dataset is presented in Section \ref{sec:factor}. Finally, conclusions and future work have been outlined in
Section \ref{sec:conclusion}.

\section{Related Work and Our Contributions}
\label{sec:relWork}
The study on impact factor~\citep{garfield} is among the most significant
works, carried to estimate the quality of publications. Before the
introduction of impact factor, several other metrics were used to
quantify the quality of research documents. These include the number
of papers published by the author over $n$ years, the number of
citations for each paper, the journals where the papers were
published, etc. Impact factors are calculated yearly starting from
1975 for those journals that are indexed in the Journal Citation
Reports\footnote{http://en.wikipedia.org/wiki/Impact\_factor.}.
Impact factor is a measure reflecting the average number of citations
to recent articles published in the journal. In Nov 2005, Jorge
E. Hirsch, came up with a new measure known as h-index, that
attempts to measure both the productivity and impact of the published
work of a researcher~\citep{hirsch}. Egghe proposed a new metric known as g-index 
to overcome the limitations of h-index~\citep{egghe2006improvement}. 
Research has also been done to find certain features that may directly or indirectly affect the impact of
a publication. \cite{DBLP:journals/corr/abs-1205-1143} proposed venue 
recommendation algorithm based on direction aware random walk with restart. \cite{wang2013} used parameters from the citation
history of a paper to quantify long-term scientific impact (total
number of citations during its lifetime). Citeseer(X), Google Scholar
and Microsoft Academic Search are academic search engines that provide
venue ranking along with basic search facilities. While these portals use
different algorithms to rank the publication venues, citation
count is the dominant feature~\citep{garfield,chakrabortyautomatic,BeelJG,conf/jcdl/YanL07}.
Several other citation-based measures have been proposed to rank the quality of documents
retrieved from a digital library \citep{Larsen:2006}, and to measure the quality 
of a small set of conferences and journals \citep{Rahm:2005:b}.

Apart from quality estimation of publications and venues, work has
also been done in venue recommendation. Studies on recommending
appropriate publication venues to the researcher for their research
paper have explored author's network of related
co-authors~\citep{luong2012,xia2013} as well as topic and writing-style
information~\citep{yang2012}. A study on identification of motivating factors for
  publication venue selection suggests that publication quality is the
  most important aspect \citep{warlick2007}. Even free public
  availability and increased exposure do not provide that strong
  incentive. A similar study \citep{West2012} suggests that the prestige of a venue depends
  on several factors such as sponsorship by national or international
  professional organization, reputation of publisher, editor and
  editorial board popularity etc. A study by \cite{zhuang2007measuring} claimed that 
  the quality of a conference is closely correlated with that of its program committee (PC). Another study on
  wellness of software engineering conferences uses features like
  author and PC stability, openness to new
  authors, inbreeding, representativeness of the PC with respect to
  the authors' community, availability of PC candidates, and
  scientific prestige \citep{Vasilescu2014251}. \cite{Souto:2007} 
  presented ``OntoQualis'', an ontological approach for domain analysis and ontology 
  prototyping aiming to classify Scientific Conferences. \cite{Franz} 
  surveyed the state-of-the-art in interpreting studies in search of conceptual and 
  methodological tools for the empirical study and assessment of conference quality.  
  \cite{Martins:2009:LAQ:1555400.1555431} presented a thorough 
  review on different features used for venue classification in past research. A similar study proposed 
  fuzzy inference models and a set of factors to access the overall quality~\citep{hussain2008ranking}. 
  They also proposed a dimensionless index called Fuzzy Index (FI) to shuffle the previously ranked research bodies. \cite{Huang2016329} observed positive correlation between impact factor and article number in scholarly journals. 
  High impact journals publish more articles. \cite{Dunaiski2016392} 
  evaluated the author and paper ranking algorithms by using a test data set of papers and authors that won renowned prizes at numerous computer science conferences.
  They observed that for ranking important papers or identifying high-impact authors, algorithms based on PageRank perform better.
   
The subject of diversity has been well studied in informetrics and scientometrics in recent years. 
Diversity of references cited by a paper logically seems to be the best gauge of intellectual integration~\citep{porter2007measuring}. 
Papers that cite more discrete subject categories pertaining to some unspecified mix of substantive topics, methods, and/or concepts are presumed to
be more interdisciplinary. Similar study by \cite{rafols2010diversity} propose a conceptual framework to capture interdisciplinarity in 
 knowledge integration, by exploring the concepts of diversity. They suggest that disciplinary diversity indicates the 
 large-scale breadth of the knowledge base of a publication. They describe diversity as heterogeneity indicators of a bibliometric set
 viewed from predefined categories. \cite{Leydesdorff201187} investigated network indicators (betweenness centrality), 
 unevenness indicators (Shannon entropy, the Gini coefficient) and Rao-Stirling measures to understand interdisciplinarity and popularity of journals.

Our study tries to look closely into the temporal fluctuations in
the underlying parameters apart from the citation counts and show
that the top-tier conferences are much more stable than the non
 top-tier ones. While addressing the problem of \textit{categorization
  of conferences into top-tier or non top-tier}, this paper tries to
answer some of the very pertinent questions:

\textit{\textbf{Question 1.} What are some of the underlying
  features behind the prestige of conferences?}

\textit{\textbf{Question 2.} How can these features be meaningfully
  used to predict the category of a given conference?}

While attempting to answer these questions, this paper makes the
following contributions. First, we put forward the hypothesis that
 top-tier conferences are much more stable than the non top-tier
ones, in terms of maintaining the same sort of diversity over the
years. We identify nine different underlying features and empirically
validate this hypothesis over a set of 110 conferences. Next, we employ
these features in an SVM model to propose a classification framework,
which we believe can act as a conflict resolution assistant for the experts.

\section{Analysis of conference-level data}
\label{sec:motivation}
A common belief in the research community is that researchers are confident about the category of the conference in the 
area of their expertise. We perform four small experiments to refute this intuition. Specifically, we  looked at 110 popular 
conferences listed in the Wikipedia entry for list of computer science conferences~\citep{wikipedia}. These 110 conferences cover
22 sub-fields of the computer science domain. Next, we present the details of our experiments and the motivating outcomes.

\subsection{Experiment I}
\label{sec:conf_cat}
In the first experiment, we compared the categories from different categorization systems. Specifically, we consider 
four state-of-the-art systems that provide conference categorizations 
and compile categories for each of the 110 conferences. Each system uses a different category label. For example, 
System 1\footnote{http://www.ntu.edu.sg/home/assourav/crank.htm.} uses the
labels: rank1, rank2, rank3 and others, System 2\footnote{http://webdocs.cs.ualberta.ca/$\sim$zaiane/htmldocs/ConfRanking.html.}
uses top-tier, second-tier and third-tier as labels, System 3\footnote{http://perso.crans.org/$\sim$genest/conf.html.} only provides top 
conferences while System 4\footnote{http://103.1.187.206/core/.} provides categories as A*, A, B, C and
unranked. For each of these systems, we labeled our conference
set such that a conference was labeled top-tier if it was listed in
the top category for all the systems (i.e., rank1 from System 1, top-tier from System 2,
any conference listed by System 3 and A* from System 4), otherwise it was labeled as
non top-tier. 

This labeling task results in four different lists. A conference is
eligible for consideration, if it is present in at least three lists. 80
out of the 110 conferences satisfied this criteria. Out of these 80
eligible conferences, we call a conference as non-conflicting (NC) if
it has been labeled using the same category in at least three lists,
otherwise it is called a conflicting conference (CC). Overall, the set
NC contains 53 conferences with 32 labeled as top-tier and 21 labeled
as non top-tier. In the rest of the paper, we call this categorized dataset as the Benchmark Dataset.

\subsection{Experiment II}
\label{sec:survey}
In the second experiment, we conduct an online survey among researchers. 28 researchers working in the field of
computer science participated in this survey. The research fields of the participants 
include Machine Learning (ML), Natural Language Processing (NLP), Complex Networks (CPN), Image Processing (IP), 
Computer Networks (CMN), Data Mining (DM), Formal Methods (FM), Computer Architecture (CA) and Information Retrieval (IR). 
While participating in the survey, each subject is shown one page per conference present in the set
CC. Within each page, the subject is shown the name of a conference,
and is asked to choose among the two categories, top-tier or non
top-tier. The subject can skip a page if he is not sure about the category of a conference. 

Considering the majority voting for each conference, we classify them into two classes: top-tier (TT) or non top-tier (NTT). 
Interestingly, only one conference FCCM has received all votes as NTT. Table \ref{tab:cat-survey-agree} compares conference categories
from experiment I and experiment II. Only 15 out of 27 conferences have the same category in both experiments. This observation further refutes the initial intuition. 

\begin{table*}[!thb]
 \caption{Comparison between experiment I and II: First column
   lists conference names. Second and third columns present major conference category in experiment I 
 and II respectively. Last column
 shows agreement between experiment I and II categories. Here, tie in the second column indicates that a conference has received equal votes in both categories in experiment I.}
 \resizebox{.85\textwidth}{!}{\begin{minipage}{\textwidth}
 \begin{tabular}{|l|c|c|c|}
  \hline
Conference Name&\parbox[t]{2cm}{Experiment I category} &\parbox[t]{2cm}{Experiment II category}&Agreement   \\ 
 \hline 
ACM Symposium on Parallel Algorithms and Architectures - SPAA&Tie&TT&NA   \\ 
 \hline 
ACM-IEEE Joint Conference on Digital Libraries - JCDL&NTT&TT&No   \\ 
 \hline 
Applications of Natural Language to Data Bases - NLDB&NTT&NTT&Yes   \\ 
 \hline 
Colloquium on Structural Information and Communication Complexity - SIROCCO&NTT&NTT&Yes   \\ 
 \hline 
Compiler Construction - CC&Tie&NTT&NA   \\ 
 \hline 
Data Compression Conference - DCC&TT&NTT&No   \\ 
 \hline 
Design Automation Conference - DAC&TT&TT&Yes   \\ 
 \hline 
Design, Automation, and Test in Europe - DATE&NTT&NTT&Yes   \\ 
 \hline 
European Conference on Object-Oriented Programming - ECOOP&NTT&NTT&Yes   \\ 
 \hline 
European Symposium on Programming - ESOP&NTT&NTT&Yes   \\ 
 \hline 
Fast Software Encryption - FSE&TT&NTT&No   \\ 
 \hline 
Field-Programmable Custom Computing Machines - FCCM&NTT&NTT&Yes   \\ 
 \hline 
Foundations of Software Science and Computation Structure - FoSSaCS&NTT&NTT&Yes   \\ 
 \hline 
International Colloquium on Automata, Languages and Programming - ICALP&Tie&NTT&NA   \\ 
 \hline 
International Conference on Computer Aided Design - ICCAD&TT&TT&Yes   \\ 
 \hline 
International Conference on Distributed Computing Systems - ICDCS&Tie&TT&NA   \\ 
 \hline 
International Conference on Information and Knowledge Management - CIKM&Tie&TT&NA   \\ 
 \hline 
International Conference on Network Protocols - ICNP&TT&NTT&No   \\ 
 \hline 
International Conference on Parallel Processing - ICPP&Tie&TT&NA   \\ 
 \hline 
International Conference on Robotics and Automation - ICRA&NTT&TT&No   \\ 
 \hline 
International Symposium on Algorithms and Computation - ISAAC&NTT&NTT&Yes   \\ 
 \hline 
Mathematical Foundations of Computer Science - MFCS&NTT&NTT&Yes   \\ 
 \hline 
Network and Operating System Support for Digital Audio and Video - NOSSDAV&NTT&NTT&Yes   \\ 
 \hline 
Principles and Practice of Constraint Programming - CP&NTT&TT&No   \\ 
 \hline 
Symposium on Graph Drawing - GD&NTT&NTT&Yes   \\ 
 \hline 
Symposium on Theoretical Aspects of Computer Science - STACS&NTT&NTT&Yes   \\ 
 \hline 
Theory and Application of Cryptographic Techniques - EUROCRYPT&TT&TT&Yes   \\ 
 \hline 
   \end{tabular}
\end{minipage}}
\label{tab:cat-survey-agree}
\end{table*}
We compute inter annotator agreement among researchers participated in experiment II using Kohen's kappa in case of agreement (as well as disagreement) between experiment I and II, separately. In case of agreement, we observe high kappa value among researchers ($\kappa=0.23$), in comparison to disagreement when kappa value was much smaller ($\kappa=0.052$).  
 These experiments illustrate that while there are many 
clear cases where expert agreement can be almost immediately achieved as to whether 
a conference is a top-tier or not, there are equally many cases that can result in a conflict 
even among the experts.

\subsection{Experiment III}
\label{sec:impact_factor}
\begin{figure*}[!thb]
  \centering
  \resizebox{0.8\linewidth}{!}{\includegraphics{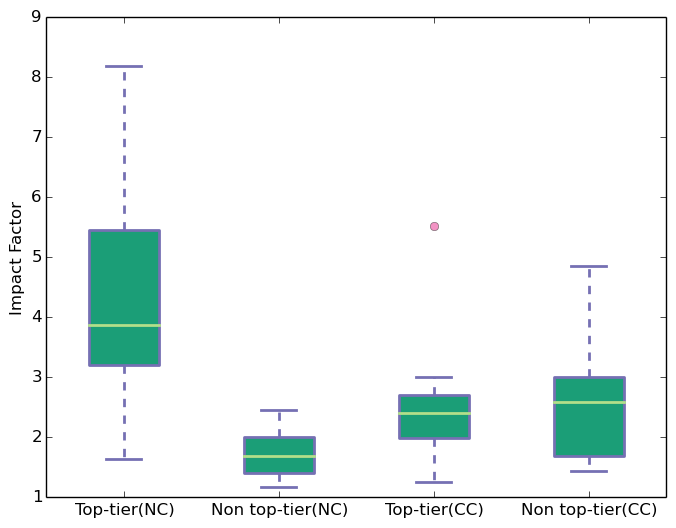}}
  \caption{Experiment III (Impact factor analysis): Majority of the top-tier conferences in set NC have high impact factor (>$3$).
  Similarly, majority of the non top-tier conferences in set NC have low impact factor (<$2$). Conferences in CC show confusing trends. 
  Majority of top-tier conferences in CC have low impact factor. Similarly, majority of non top-tier conferences in CC have high impact factor.} \label{impact_factor}
\end{figure*}

The impact factor ($IF$) is a standardized measure created by the Institute of Scientific Information (ISI) which can be used to measure the way a journal
receives citations to its articles over time. It is calculated by dividing the number of current citations a journal receives to articles published 
in the two previous years by the number of articles published in those same years. So, for example, the 2010 $IF$ is the citations in 2010 to
articles published in 2009 and 2008 divided by the number of articles published in 2009 and 2008. $IF$ concerns only journals included in Thomson Reuters Journal Citation Reports. While there are no $IF$s reported for conference proceedings, it can be computed based on the standard definition and there is a common consensus in the research community that the $IF$ of a conference should be directly proportional to its scientific impact. Figure \ref{impact_factor} presents $IF$ values for
NC and CC conferences. For conferences in the set NC, we observe a clear demarcation between top-tier and non top-tier conferences' $IF$ values. 
Majority of the top-tier conferences (25 out of 32) in NC have $IF>3$. Similarly, majority of the non top-tier conferences (15 out of 21) have $IF<2$. 
Further, we consider the category of each conference in CC based on expert decision described in Section \ref{sec:survey}. 
Interestingly, in this case, we find no relation between impact factor and conference category. 
In fact, majority of conferences (11 out of 17) marked as non top-tier have $IF>2$. Similarly, only two out of ten top-tier conferences have $IF>3$. This experiment further illustrates that for the conflicting category, $IF$ may not be a good separator. 

\subsection{Experiment IV}
\label{sec:acceptance_rate}
\begin{figure*}[!thb]
  \centering
  \resizebox{1\linewidth}{!}{\includegraphics{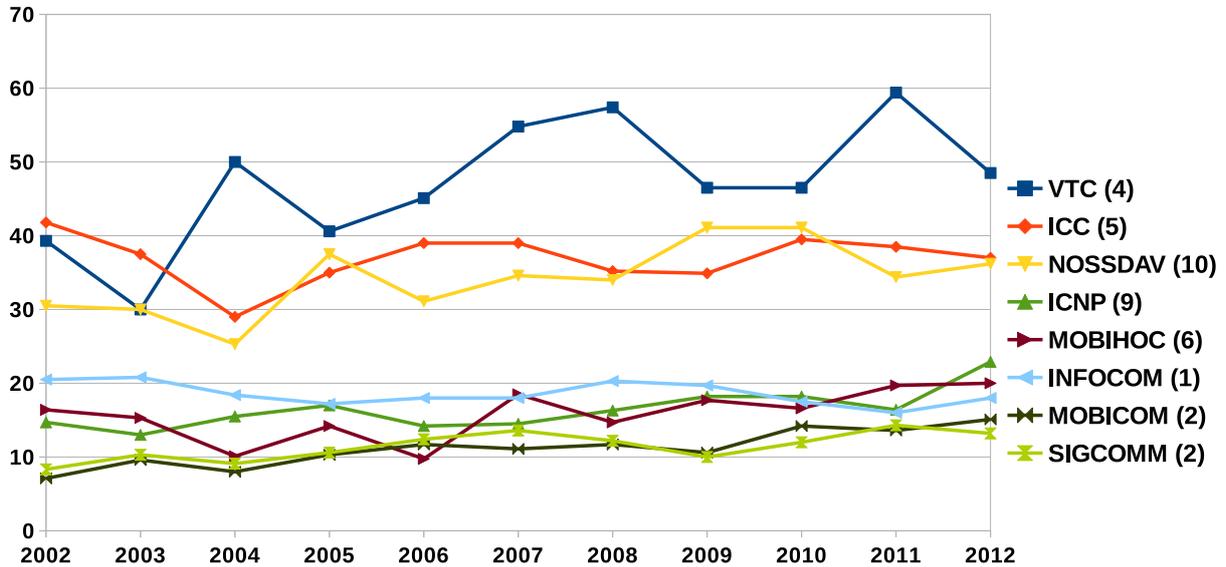}}
  \caption{Acceptance rate for the top ten computer networks conferences over the years. Number inside brackets represents rank of the conference assigned by Microsoft academic search. Two conferences in top five, namely, ICC and VTC have high acceptance rate (\textasciitilde30). VTC (rank=4) has significantly higher acceptance rate ($>37\%$) than 
NOSSDAV (rank=10). Similarly, acceptance rate of ICNP (rank=9) is significantly low (\textasciitilde15). Note that two conferences (IPSN and SenSys) are not present due 
to unavailability of data.} \label{acceptance_rate}
\end{figure*}

It is also generally believed that the acceptance rate of a conference is inversely proportional to its 
scientific impact~\citep{Martins:2009:LAQ:1555400.1555431}.
\cite{Vasilescu2014251} observed strong negative linear correlation, 
suggesting that conferences with higher acceptance rates indeed have lower scientific impact.
In the fourth experiment, we conduct similar study on computer networks conferences. We select top ten computer 
networks conferences from Microsoft academic search \footnote{http://academic.research.microsoft.com/RankList?entitytype=3\&topDomainID=2\&subDomainID=14\&last=0\&start=1\&end=100.}. 
In order to understand the relation between acceptance rate and conference tier, we collected acceptance rate statistics for the above conferences\footnote{https://www.cs.ucsb.edu/~almeroth/conf/stats/}.
Figure \ref{acceptance_rate} presents temporal acceptance rates for top ten computer networks conferences.
Note that Figure \ref{acceptance_rate} shows eight conferences instead of ten due to unavailability of data between the year 2002-2012.

Two conferences in top five namely ICC and VTC have high acceptance rate (\textasciitilde30). VTC (rank=4) has significantly higher acceptance rate ($>37\%$) than 
NOSSDAV (rank=10). Similarly, acceptance rate of ICNP (rank=9) is significantly low (\textasciitilde15). 
We therefore observe that, it is not always true that all the top-tier conferences have low acceptance rate, and non top-tier conferences have high acceptance rate. There are many cases where clear demarcation of acceptance rate between top-tier and non top-tier is not found. In the current work, we do not explore effect of acceptance rate on conference popularity, due to unavailability of temporal acceptance rate statistics 
for majority of the conferences. 

Motivated with these experiments, we aim to propose a system that can assist
domain experts in deciding the category of the conferences. This system presents a comparative 
overview of the temporal profile of queried conference with the well established 
top-tier/non top-tier conferences. \textit{ConfAssist} tries to serve as an aid in such cases by 
increasing the confidence of the experts in their decision.

\section{Dataset}
\label{sec:dataset}
This paper uses a pre-processed dataset, crawled from Microsoft Academic Search (MAS)\footnote{http://academic.research.microsoft.com}
in the year\\ 2013~\citep{chakrabortytowards}.
\subsection{Dataset Description}
In this dataset, each paper is associated with various bibliographic information -- the title of the paper, 
a unique index for the paper, its author(s), the affiliation of the author(s),
the year of publication, the publication venue, the related field(s)\footnote{Note that, the different 
sub-branches like Algorithms, AI, Operating Systems etc. constitute different ``fields'' of the computer science domain.
} of the paper, the abstract and the keywords of the papers. All author names 
are disambiguated by MAS itself using a unique identifier. For our study, we consider papers published from 1999 to 2010 for 
110 conferences described in Section \ref{sec:motivation}.
The main criterion behind the choice of these 110 conferences was the availability of yearly data during
1999-2010. Table \ref{tab:dataset} details various statistics for
the full dataset as well as the filtered dataset for the selected 110 conferences.\\

\begin{table}[!thb]
 \centering
 \caption{ General comparison between statistics of complete and filtered dataset.}\label{dataset}
  \begin{tabular}{|l|l|l|}
  \hline
  &Complete&Filtered\\
   \hline
  Year range&1970-2013&1999-2010\\
   \hline 
  Number of unique venues&6,143&110\\
   \hline
  Number of computer science fields&24&22\\
   \hline
  Number of publications&2,473,171&113,425\\
   \hline
  Number of authors&1,186,412&138,923\\
   \hline
 Avg. number of papers per author &5.18&5.22\\
   \hline
 Avg. number of authors per paper  &2.49&2.35\\
\hline
  \end{tabular}
\label{tab:dataset}
\end{table}

\subsection{Curation of Dataset}
Crawling the papers of computer science domain present in MAS was started on 
March 2014 and took six weeks to complete. The automated crawler initially used the rank-list  
given by MAS for each field to obtain the list of unique paper IDs. The paper IDs were then used 
to fetch the metadata of  the publications. \cite{chakrabortytowards} used Tor\footnote{http://torproject.org.in/} 
different systems in order to avoid overloading a particular server with bursty traffic. 
They employed random exponential back-off time whenever the server or the 
connection returned some error and sent the request again. They followed  the robot restrictions imposed by the servers to 
ensure efficient crawling of data from both the client and the server perspective. The completely crawled 
dataset contained all the information related to around 2.5 million papers which are further distributed over 24 fields
domain as shown in Table \ref{tab:dataset}.

\subsection{Preprocessing of the curated dataset}
The crawled data had several inconsistencies that were 
removed through a series of steps. First, few forward citations were removed which point to the papers published after 
the publication of the source paper. These forward citations 
appear because there are certain papers that are initially uploaded in public repositories (such as http://arxiv.org/) 
but accepted later in a publication venue. Further, they considered only those papers published in between 1970 and
2010 because this time period seemed to be most consistent since most of the articles published at that time period
are available in the dataset. Only those papers are considered
that cite or are cited by at least one paper (i.e., isolated nodes with zero in-degree and zero out-degree have been removed). An 
advantage of using this dataset is that the problem arising 
due to the ambiguity of named-entities (authors and publication venues) has been completely resolved by MAS itself,
and a unique identity has been associated with each author, 
paper and publication venue. Some of the authors were 
found missing in the information of the corresponding papers which were resolved by the DOI (Digital Object Identifier) 
of the publications. We double checked the filtered
papers having the author 
and metadata information from DOI and kept only the consistent ones. Some of the references that pointed to such
papers absent in the dataset (i.e., dangling references) were
also removed.

\section{Features}
\label{sec:features}
The main emphasis of the current study is to find the underlying
parameters, specifically those that can give useful insights into the
difference in the dynamics of the top-tier and the other conferences over
the years. We mainly focus on the features that indicate the diversity
pattern of the conferences. We select nine different features and study
the dynamics of the conferences in terms of how these parameters
change over the years. Next, we discuss these features in detail. The
features have been grouped in two main categories; features based on
diversity pattern in the accepted papers and features based on the
co-authorship network of authors of the accepted papers.

\subsection{Features based on diversity pattern in the accepted
  papers}
We identified certain features based on how diverse the accepted
papers are, how diverse the publication age of the authors are and
the proportion of new authors. These features have been
described in detail below. 

\subsubsection{Conference Reference Diversity Index (CRDI)}
The first feature we study is CRDI, which is related to the reference
diversity patterns. Reference diversity measures how diversified
are the fields, that have been referred to by a publication. We make use
of the fact that MAS has mapped each publication to a predefined set
of sub-fields under the domain of Computer Science. \cite{chakrabortytowards} proposed the Reference
Diversity Index (RDI), where they use this sub-division to measure the
reference diversity. In this paper, we extend this definition to measure the reference diversity of a
conference for a particular year. We define the CRDI metric as \\
\begin{equation}
CRDI(c,i) = - \sum_{n=1}^{N} p_{n}\log_2p_{n} 
\end{equation}
\begin{equation}
p_{n} = {\frac {f_{c,i,n}}{t_{c,i}}}
\end{equation}
\begin{equation}\label{CRDI-def}
\Delta CRDI^c_{(i,i+1)} = | CRDI(c,i) - CRDI(c,i+1) |
\end{equation}
where $CRDI(c,i)$ denotes the $CRDI$ value for the conference $c$ in
the $i^{th}$ year, $f_{c,i,n}$ denotes the number of
papers tagged with the $n^{th}$ sub-field for the conference-year pair $(c,i)$ and
$t_{c,i}$ denotes the number of publications for the pair $(c,i)$. A
high $CRDI$ indicates that the conference is highly diverse (or
inter-disciplinary). Difference of $CRDI$ values $\Delta
CRDI^c_{(i,i+1)}$ for consecutive years plays a very significant role;
in that, if this value is small, it indicates that the conference is
 stable in terms of the amount of diversified references in the
research papers published in the conference.

\subsubsection{Conference Keyword Diversity Index (CKDI)}
CKDI is the second feature that we study, which is related to the
keyword diversity pattern of a conference. Similar to the field
mapping, MAS has also mapped each publication to a global set of
keywords. The mapped keywords are extracted on the basis of
publication abstract and keywords. \cite{chakrabortytowards}
proposed the Keyword Diversity Index (KDI) to represent the diversity
in the paper keywords. In this paper, we extend this definition to
measure the keyword diversity of a conference in a particular year. We
define the CKDI metric along the similar lines as that of the CRDI metric: \\
\begin{equation}
CKDI(c,i) = - \sum_{n=1}^{K_{c,i}} p_{n}\log_2p_{n} 
\end{equation}
\begin{equation}
p_{n} = {\frac {k_{c,i,n}}{tk_{c,i}}}
\end{equation}
\begin{equation}
\Delta CKDI^c_{(i,i+1)} = | CKDI(c,i) - CKDI(c,i+1) |
\end{equation}
where $CKDI(c,i)$ denotes the $CKDI$ value for the conference $c$ in
the $i^{th}$ year, $K_{c,i}$ denotes the number of unique
keywords, $k_{c,i,n}$ denotes the count of $n^{th}$ keyword for the
pair $(c,i)$ and $tk_{c,i}$ denotes the total count of all keywords
for the pair $(c,i)$. Similar to CRDI, a high CKDI value indicates
that the keywords used in the conference papers are diverse for
that year. Similarly, if the consecutive year differences in $CKDI$
are relatively small, it signifies that conference papers are stable
in terms of the topic diversity of the research papers, published in the conference.

\subsubsection{Conference Author Diversity Index (CADI)}
CADI is the third feature that we study, which corresponds to the
fraction of authors with diversified research interests publish in a
conference. Author Diversity Index (ADI) corresponds to how diverse are the author's
publications in the last five years \citep{chakrabortytowards}. If $A_{c,i}$ denotes the total number
of authors in a conference \textit{c} in the $i^{th}$ year, this index
is calculated over all the $A_{c,i}$ authors 
and the Conference Author Diversity Index (CADI) is expressed as the average ADI value of these $A_{c,i}$ authors.
\begin{equation}
ADI(i,j) = - \sum_{n=1}^N p_{n}\log_2p_{n} 
\end{equation}
\begin{equation}
p_{n} = {\frac {a_{i,j,n}}{t_{i,j}}}
\end{equation}
\begin{equation}
CADI(c,i) = {\frac {\sum_{\forall j \in A{c,i}}  ADI(j,i) }{A_{c,i}}} 
\end{equation}
\begin{equation}
\Delta CADI^c_{(j,j+1)} = | CADI(c,j) - CADI(c,j+1) |
\end{equation}
where $ADI(i,j)$ denotes the ADI value for the $i^{th}$ author in the
$j^{th}$ year, $N$ denotes the number of unique sub-fields, $a_{i,j,n}$
denotes the number of papers published by the author $i$ in the
$n^{th}$ sub-field during the last five years, $j-4$ to $j$ and $t_{i,j}$
denotes the total number of papers published by the author $i$ in the
last five years. $\Delta CADI^c_{(j,j+1)}$ represents the CADI difference
between the consecutive years $(j,j+1)$ and is similar to the previous
two measures; a low difference signifies that the conference is stable
in terms of the diversity of the authors, who are publishing in the conference.

\subsubsection{Proportion of New Authors (PNA)}
The fourth feature that we use is the proportion of new authors in a
conference. The main intuition behind using this feature is to
explore whether the fraction of papers with new authors is roughly the
same over the years for a top-tier conference. A paper in a particular conference contributes to this
proportion if all of the authors of the paper are new, i.e., none of
the authors have any publication in this conference in the last five
years.

Thus, Conference New Author count ($CNA_{c,i}$) is calculated for the
conference $c$ for the $i^{th}$ year using the above
definition. $PNA(c,i)$ and $\Delta PNA^c_{(i,i+1)}$ are computed as follows:
\begin{equation}
PNA(c,i) =  \frac {CNA_{c,i}}{n_{c,i}}
\end{equation}
\begin{equation}
\Delta PNA^c_{(i,i+1)} = | PNA(c,i) - PNA(c,i+1) |
\end{equation}
where $n_{c,i}$ denotes the total number of unique authors for the pair
$(c,i)$. $\Delta PNA^c_{(j,j+1)}$ represents the difference in PNA
values for consecutive years $(j,j+1)$. A low difference signifies
that the conference is stable in terms of what proportion of the
authors are new in a given year.

\subsubsection{Conference Author Publication Age Diversity Index (CAAI)}
This feature is related to the extent of diversity of the publication experience of the authors in a
conference. The prime intuition behind using this feature is to study
whether the top-tier conferences have more inclination towards
maintaining similar publication-age diversity (or diversity in
terms of publication experience of the authors) over time. Author publication age is calculated from
her first publication year in the entire MAS dataset. Let $F_j$ denote the first publication
year of the $j^{th}$ author. We define the CAAI metric as: \\
 \begin{equation}
   AA(i,j)=\begin{cases}
     i-F_j, & \text{if $F_j<i$}.\\
     0, & \text{otherwise}.
   \end{cases}
 \end{equation}
\begin{equation}
CAAI(c,i) = - {\sum_{\forall j \in AC{c,i}}} p_{n}\log_2p_{n} 
\end{equation}
\begin{equation}
p_{n} = {\frac {a_{c,i,n}}{t_{c,i}}}
\end{equation}
\begin{equation}
\Delta CAAI^c_{(i,i+1)} = | CAAI(c,i) - CAAI(c,i+1) |
\end{equation}
where $AA(i,j)$ represents $j^{th}$ author's age in the year $i$. Set
$AC{c,i}$ contains all unique AA values for the conference $c$ in the $i^{th}$ year;
$t_{c,i}$ denotes the total number of authors in the confidence $c$ for the $i^{th}$ year and $a_{c,i,n}$ denotes the number
of authors who have publication age equivalent to $n$ for the conference $c$ in the year $i$. 

\subsection{Co-authorship Network Features}
Next, we study features related to the co-authorship behavior of the
authors in a conference. We build the co-authorship network for
each year using the complete MAS dataset. Further, we extract induced sub-graph from this reference
network for each conference and for each year. We aim to capture fluctuations in the
network properties of these conferences over the 12-year time
period. 

\textbf{Co-authorship network description:}
We build co-authorship network from the author information present in the paper metadata. 
For a particular year $y$, we consider all the publications from the year $1971$ to $y-1$ to
build this network ($G(V,E,y)$). Here, the nodes($V$) correspond to the
authors and an edge($E$) between two nodes is weighted as per their co-authorship
count, i.e., higher the edge weight, higher the co-authorship count. 

\textbf{Induced co-authorship sub-graph description:}
We extract induced sub-graph ($g(v,e,y,c)$) from $G(V,E,y)$ for each conference ($c$) and for each year ($y$).
Here, the nodes ($v$) correspond to the authors present in the $c^{th}$ conference in the $y^{th}$ year and edges ($e$) correspond to co-authorship
count between two nodes. Given a vertex set $v$, we choose edges $e$ from $E$ to create graph $g$, if both the endpoints are present in $v$. 
Note that $v \subset V$ and $e \subset E$. 

Next, we define four features based on the induced co-authorship network. 

\subsubsection{Degree Diversity Index (DDI)}
This feature corresponds to the diversity in the degree of co-authorship of conference
authors. The degree of a node corresponds to the sum of the edge-weights incident on that node. 
The intuition behind this feature is to understand the
fluctuations in the overall collaborative behavior of the authors in the conference assuming that the 
co-authorship behavior is a close representative of the collaborative behavior of the authors. For
conference \textit{c} in the $i^{th}$ year, we define DDI as:
\begin{equation}
DDI(c,i) = - {\sum_{\forall n \in D{c,i}}} p_{n}\log_2p_{n} 
\end{equation}
\begin{equation}
p_{n} = {\frac {d_{c,i,n}}{td_{c,i}}}
\end{equation}
\begin{equation}
\Delta DDI^c_{(i,i+1)} = | DDI(c,i) - DDI(c,i+1) |
\end{equation}

where set $D_{c,i}$ contains the number of unique degrees, $d_{c,i,n}$ 
 denotes the number of nodes with degree $n$ and $td_{c,i}$ denotes 
the total degree of all authors for the pair $(c,i)$. If the consecutive year differences 
for the $DDI$ values are relatively small, it signifies that the conference
promotes similar extent of collaboration among the authors over the years.
On the other hand, conferences having 
high consecutive year differences in $DDI$ are still experimenting with the
trade-off between high and low collaborative authors.

\subsubsection{Edge Strength Diversity Index (EDI)}
This feature corresponds to the diversity in the edge weights of
conference authors. The intuition behind this feature is to understand
the fluctuations in the choice of the co-authors for a given author in a conference. For
conference \textit{c} in the $i^{th}$ year, we define EDI as:
\begin{equation}
EDI(c,i) = - {\sum_{\forall n \in E{c,i}}} p_{n}\log_2p_{n}
\end{equation}
\begin{equation}
p_{n} = {\frac {e_{c,i,n}}{te_{c,i}}}
\end{equation}
\begin{equation}
\Delta EDI^c_{(i,i+1)} = | EDI(c,i) - EDI(c,i+1) |
\end{equation}

where set $E_{c,i}$ contains the number of unique edge weights, $e_{c,i,n}$ denotes the 
number of edges with weight $n$ and $te_{c,i}$ denotes the sum of all edge weights for the pair $(c,i)$. 

\subsubsection{Average Closeness centrality (ACC)}
Closeness centrality is defined as the inverse of farness, which in turn, is the sum of distances to all other nodes.
It measures how close a node is to all other vertices in the graph. In
the co-authorship network, it represents 
closeness of a author to other authors in terms of collaboration. The motivation behind this
feature is to study the changes in the most central nodes over time for a conference.
For the conference \textit{c} in the $i^{th}$
year, $ACC$ values are computed as follows:
\vspace{-0.2cm}
\begin{equation}
ACC(c,i) =  \frac {{\sum_{\forall n \in A{c,i}} CC_n}}{n_{c,i}}
\end{equation}
\begin{equation}
\Delta ACC^c_{(i,i+1)} = | ACC(c,i) - ACC(c,i+1) |
\end{equation}
where $A_{c,i}$ denotes the total number of authors, $n_{c,i}$ denotes
the total number of unique authors having non-zero closeness centrality and $CC_n$ represents the closeness value of $n^{th}$ author
for the pair $(c,i)$. The higher the value of $ACC$, the more compact is a conference community.

\subsubsection{Average Betweenness Centrality (ABC)}
Betweenness centrality \citep{Linton}  measures the ``importance'' of
each node in the network. For the conference \textit{c} in the
$i^{th}$ year, average betweenness centrality value can be computed
as:
\begin{equation}
ABC(c,i) =  \frac {{\sum_{\forall n \in A{c,i}} BC_n}}{n_{c,i}}
\end{equation}
\begin{equation}
\Delta ABC^c_{(i,i+1)} = | ABC(c,i) - ABC(c,i+1) |
\end{equation}
where $A_{c,i}$ denotes the total number of authors, $n_{c,i}$ denotes
the total number of unique authors having non-zero betweenness 
%PG:
%same as above
centrality and $BC_n$ represents betweenness value of $n^{th}$ author
for the pair $(c,i)$.

\section{Feature Analysis}
\label{sec:analysis}
Once we identified nine different quantities that might be helpful to
separate a top-tier conference from a non top-tier one, we tried to
study these quantities in further details. For each of the nine quantities,
we use three different parameters: the mean value, the median and the
standard deviation. For example, corresponding to $CRDI$, we
use three different features, i.e., mean over the 11 $\Delta CRDI^c_{(i,i+1)}$
values for the conference $c$, median over these values as well as the
standard deviation over these values. Thus, we get 27 features for
each conference in our dataset. For the sake of visualization, we divide features into three buckets, features 1-9, features 10-18 and features 19-27. Table \ref{tab:three buckets} presents division of features into three buckets. In this section, we do a thorough
analysis of these 27 features using the benchmark dataset, described in Section
\ref{sec:motivation}.

\begin{table}[!thb]
 \centering
 \caption{ Division of features into three buckets.}
  \begin{tabular}{|l|l||l||l||l|l|}
\hline
\multicolumn{2}{|c||}{Bucket I} &\multicolumn{2}{|c||}{Bucket II} & \multicolumn{2}{|c|}{Bucket III} \\\hline\hline
1&CRDI mean&10&PNA mean&19&EDI mean\\
\hline
2&CRDI median&11&PNA median&20&EDI median\\
\hline
3&CRDI stddev&12&PNA stddev&21&EDI stddev\\
\hline
4&CKDI mean&13&CAAI mean&22&ACC mean\\
\hline
5&CKDI median&14&CAAI median&23&ACC median\\
\hline
6&CKDI stddev&15&CAAI stddev&24&ACC stddev\\
\hline
7&CADI mean&16&DDI mean&25&ABC mean\\
\hline
8&CADI median&17&DDI median&26&ABC median\\
\hline
9&CADI stddev&18&DDI stddev&27&ABC stddev\\
\hline
  \end{tabular}
\label{tab:three buckets}
\end{table}

\subsection{Comparing top-tier and non top-tier using features}
\begin{figure}[!thb]
  \centering
  \resizebox{.7\linewidth}{!}{\includegraphics{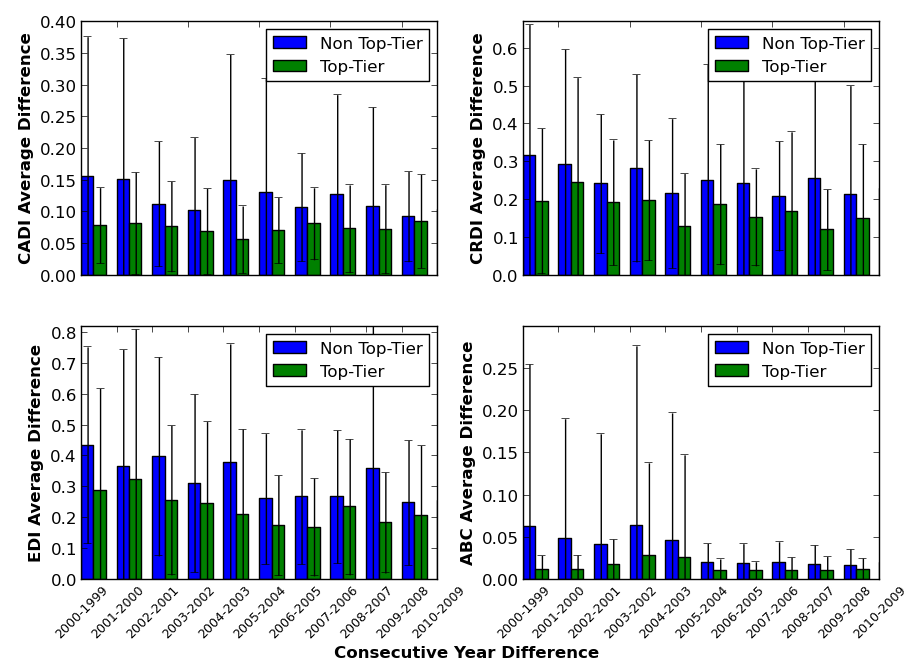}}
  \caption{Comparison between top-tier and non top-tier using $\Delta$CRDI, $\Delta$CADI, $\Delta$EDI and $\Delta$ABC's average and standard deviation profiles.  X-axis 
  denotes 11 consecutive year-differences and y-axis denotes the mean
  of the difference values across various conferences in a category, with error bars showing standard 
  deviation of the difference values.} \label{year-wise}
\end{figure}

First, we try to identify if there are differences between the feature
values, obtained for the top-tier and non top-tier conferences in
general and whether these are consistent with our hypothesis. For each
of the nine quantities, 110 conferences and categorization defined in
Section \ref{sec:motivation}, we first compute the year-wise differences for these
quantities, e.g. $\Delta ABC$ etc.
Then, we contrast the mean and the standard deviation values of the
year-wise differences for all top-tier conferences with 
all non top-tier conferences. Figure \ref{year-wise}
presents the comparison between these categories using $\Delta$CRDI,
$\Delta$CADI, $\Delta$EDI and $\Delta$ABC's average and standard
deviation profiles. X-axis denotes 11 consecutive year-differences 
and y-axis denotes the mean of the difference values with error bars showing standard deviation of the difference values. The 
corresponding values for the top-tier and non top-tier conferences are plotted using green and blue bars respectively. 
An analysis of these plots gives a clear indication that for all the
four example quantities, the blue bars are higher than the green bars,
i.e. there is a higher fluctuation for the non top-tier 
conferences (yearwise differences denoted by the height of blue bar
are higher) as compared to the top-tier conferences (yearwise
differences denoted by the green bar are lower). Table \ref{tab:mean_std_dev_tt_ntt} presents similar statistics for all the 9 features. This favors our
hypothesis that the top-tier conferences are much more stable than the
non top-tier conferences. Further, while the
standard deviation for the top-tier conferences is relatively low, it is
significantly higher for the non top-tier conferences. 

We observe that out of nine features, differences between the means of the top tier and non-top tier conferences are statistically significant with a Student t-test ($p=0.05$) for three features, namely CRDI ($p=0.024$), CKDI ($p=0.036$) and DDI ($p=0.002$). ABC, however, present a moderate trend toward significance with $p=0.077$. For other 5 features, mean differences are not statistically significant.

\begin{figure*}[!thb]
  \centering
  \resizebox{1\linewidth}{!}{\includegraphics{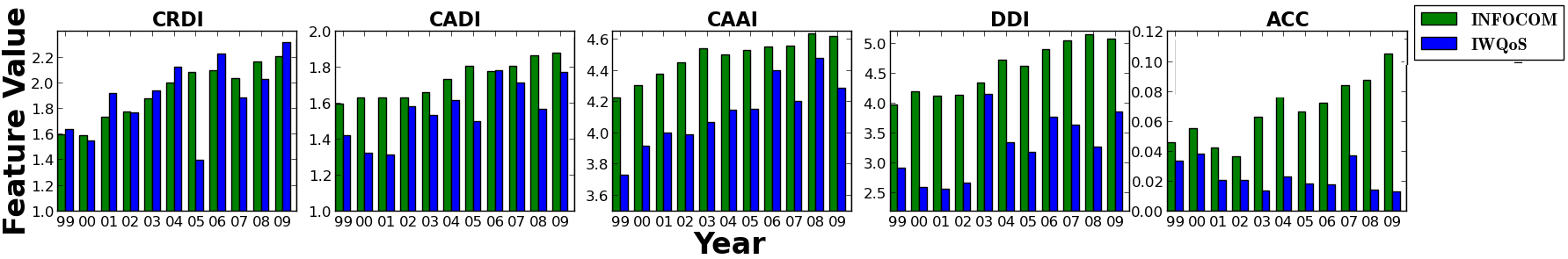}}
  \caption{Comparison between INFOCOM (top-tier) and IWQoS (non top-tier) raw feature values using CRDI, CADI, CAAI, DDI and ACC on a yearly 
scale.} \label{5_feature}
\end{figure*}

We also analyzed the raw quantities in addition to the consecutive year differences. As a representative example, Figure \ref{5_feature} 
presents comparison between INFOCOM (top-tier) and IWQoS (non top-tier) using five features, CRDI, CADI, CAAI, DDI and ACC on a yearly 
scale. One observation is that in majority of features,
values are much higher for INFOCOM than for IWQoS. However, more
importantly, and as also noted in Figure \ref{year-wise}, we again find that the differences in the 
feature values over the consecutive time periods are significantly lower for INFOCOM in contrast to IWQoS. Similar study over the entire set of conferences is reported in Table \ref{tab:raw_feature_analysis}. Table \ref{tab:raw_feature_analysis} presents statistics for conference pairs (one top tier and one non-top tier from same field of study). We represent a +ve trend if top tier conferences have higher raw values than non-top tier conferences and represent -ve trend otherwise. The first column denotes the minimum number of years for the trend. For example, if we consider minimum eight years to represent a trend, 74\% conference pairs have higher raw CAAI values for top tier than non-top tier conferences. 

This study yields some interesting observations. As noted, majority of conference pairs show significant difference in raw feature values. For six features (CADI, CAAI, DDI, EDI, ACC and ABC), top tier conferences have higher raw values than non-top tier conferences (represented by +ve). Rest three features show opposite trends (represented by -ve).

\begin{table}[!thb]
\centering
 \caption{Mean and standard deviation for top tier and non-top tier conferences averaged over 11 consecutive year differences.}
 \begin{tabular}{|l|l|l|l|l|}
\hline\multirow{2}{*}{Feature}&\multicolumn{2}{c|}{Non-top tier} &\multicolumn{2}{c|}{Top tier}\\\cline{2-5}
&Mean&Std. Dev. & Mean&Std. Dev.\\\hline
CRDI&0.25&0.259&0.173&0.178\\\hline
CKDI&0.536&0.504&0.396&0.419\\\hline
CADI&0.121&0.146&0.076&0.068\\\hline
PNA&0.1&0.077&0.097&0.068\\\hline
CAAI&0.206&0.232&0.175&0.16\\\hline
DDI&0.386&0.323&0.335&0.28\\\hline
EDI&0.323&0.292&0.23&0.265\\\hline
ACC&0.02&0.034&0.018&0.019\\\hline
ABC&0.034&0.087&0.015&0.034\\\hline
  \end{tabular}
\label{tab:mean_std_dev_tt_ntt}
\end{table}

\vspace{-0.1cm}

\begin{table}[!thb]
\centering
 \caption{Proportion of conference pairs: Majority of conference pairs (one top tier and one non-top tier from same field of study) shows significant difference in raw feature values. For six features (CADI, CAAI, DDI, EDI, ACC and ABC), top tier conferences have higher raw values than non-top tier (represented by +ve). Rest three features show opposite trends (represented by -ve). Column 1 shows minimum number of years required to represent +ve or -ve trend.}
 
 \begin{tabular}{|c|c|c|c|c|c|c|c|c|c|}
  \hline
&CRDI(-)&CKDI(-)&CADI(+)&PNA(-)&CAAI(+)&DDI(+)&EDI(+)&ACC(+)&ABC(+)\\\hline
8&0.65&0.53&0.61&0.57&0.74&0.55&0.40&0.68&0.70\\\hline
9&0.55&0.46&0.57&0.57&0.68&0.53&0.27&0.63&0.68\\\hline
10&0.40&0.46&0.53&0.46&0.57&0.42&0.17&0.53&0.61\\\hline
11&0.38&0.46&0.44&0.38&0.48&0.31&0.10&0.38&0.51\\\hline
12&0.27&0.34&0.27&0.27&0.29&0.23&0.04&0.31&0.42\\\hline
\end{tabular}
\label{tab:raw_feature_analysis}
\end{table}
\vspace{-0.2cm}

\subsection{Fieldwise comparison of representative conferences}
\begin{figure*}[!thb]
  \centering
  \resizebox{.7\linewidth}{!}{\includegraphics{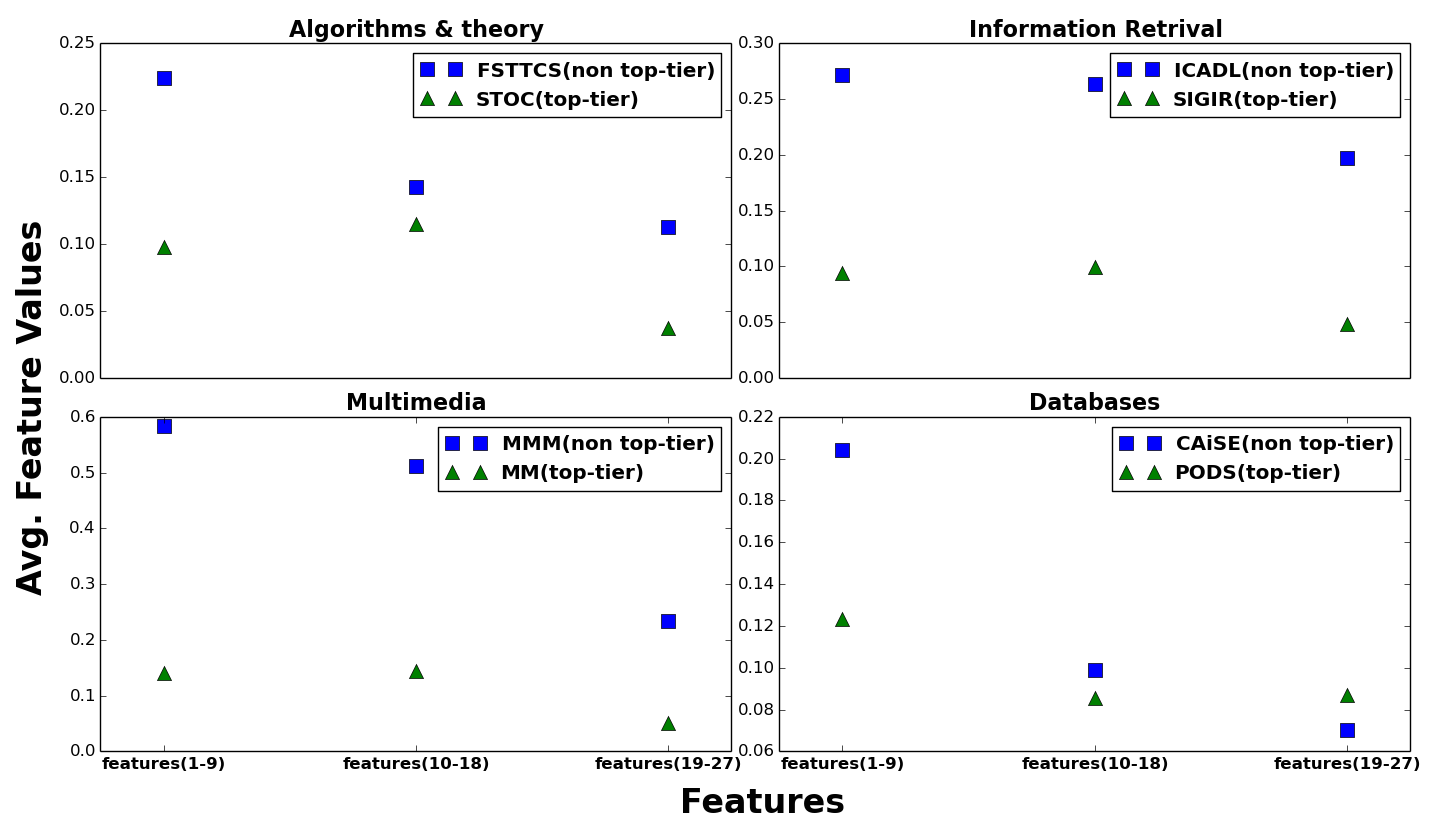}}
  \caption{Comparison of feature values of top-tier and non top-tier in four computer science fields. For each field, we consider 
  two representative conferences from each category.} \label{compare_tt_ntt}
\end{figure*}

We compare feature values of top-tier conferences with non top-tier in each field. Figure \ref{compare_tt_ntt} shows 
 plots for four computer science fields, namely, {\sl Algorithms and
   Theory}, {\sl Multimedia}, {\sl Information Retrieval} and {\sl Databases}. In each plot, 
 we consider a representative top-tier and non top-tier conference
 from these fields. In these plots, we
 compare the representative conferences for the set of 27 features. As described before in Table \ref{tab:three buckets}, we divide our features into three buckets, features 1-9, features
 10-18 and features 19-27. For each conference, we plot the average of the
 feature values within a bucket and thus, we obtain three values for
 each conference corresponding to three buckets of features.

\begin{figure*}[!thb]
  \centering
  \resizebox{.7\linewidth}{!}{\includegraphics{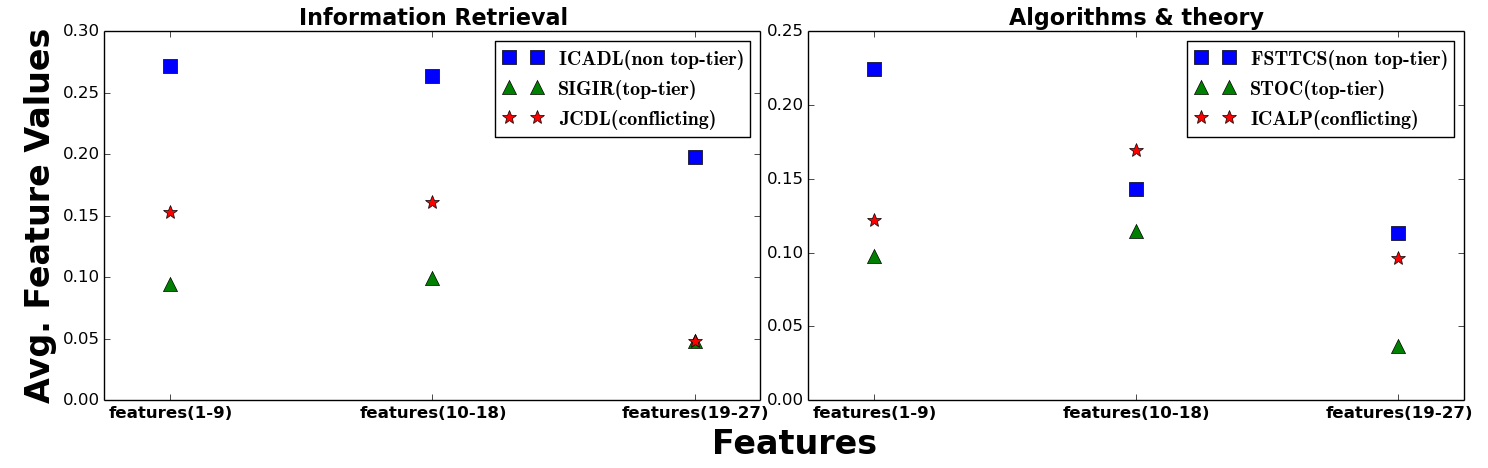}}
  \caption{Comparison of feature values of conflicting conferences with top-tier and non top-tier. We consider two representative 
  fields, Algorithms and theory and Information retrieval. Conflicting examples are taken from benchmark II.} \label{conflict}
\end{figure*}

 One straightforward observation is that the feature values for top-tier conferences
 are lower than those for the non top-tier. This
 observation holds for all of the four fields, considered in this
 figure. On further analysis, we observe that the separation between
 the two conferences is proportional to the ratio of their respective
 field ratings. We further analyze the behavior of some of the
 conflicting conferences. Figure \ref{conflict} presents plots of two
 conflicting conferences, ICALP and JCDL, compared against the
 representative conferences of their fields,
 {\sl Algorithms and Theory} and {\sl Information Retrieval}
 respectively. At least for the first two buckets, the feature values for these conferences
lie between the features values of the top-tier
 and the non top-tier conferences in their field. This is in agreement with the fact that
 such conferences are the potential sources of conflict among the different conference categorization portals.
   
\vspace{-0.1cm}
\vspace{-0.1cm}

\subsection{Comparison of newly starting conferences with top-tier and non top-tier}
\begin{figure}[!thb]
  \centering
  \resizebox{.7\linewidth}{!}{\includegraphics{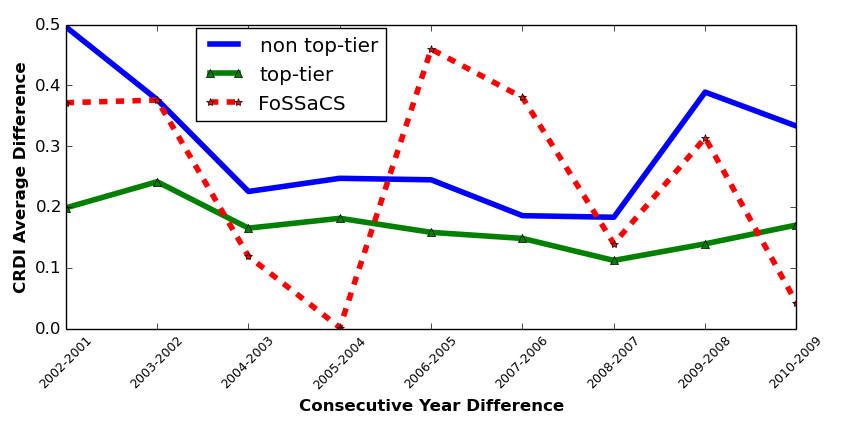}}
  \caption{Comparison of year-wise profile for FoSSaCS and average of all top-tier and non top-tier using $\Delta$CRDI.} \label{Fossacs_comparison}
\end{figure}

 We also made an attempt to compare a newly starting conference with
 top-tier and non top-tier conferences. The motivation behind this
 study was to explore, whether there are some initial signals, that can
 be used to predict future popularity of a conference. For this
 analysis, we consider \textit{International Conference on Foundations of Software Science and Computation Structures (FoSSaCS)} (started in 1999) as the representative example. Figure \ref{Fossacs_comparison}
 shows comparison of year-wise profile for FoSSaCS with
 the average values for all the top-tier and non top-tier using $\Delta$CRDI values. As noted from  Figure \ref{Fossacs_comparison}, the $\Delta$CRDI values for FoSSaCS are highly fluctuating resulting in closely matching the characteristics of non top-tier conference profiles.

\subsection{Effect of average number of publication on features}
Finally, we also wanted to analyze if the number of papers accepted in
a conference have some correlation with our feature values. We took two
representative features $\Delta$ CRDI(stddev) and $\Delta$ EDI(stddev)
to study the correlation between publication count and features. For
this study, we divide our conferences into three buckets. First
bucket consists of conferences having average publication count less 
than 35. Similarly, conferences in bucket 2 have average publication
count between 35 and 150, while the rest of the conferences are in bucket 3. Further,
we divide each bucket into two sets, top-tier and non top-tier conferences. Out of 8 conferences in bucket 1, five are top-tier and three 
are non top-tier. Similarly, bucket 2 has 34 conferences with 21 top-tier and 13 non top-tier. Five top-tier and five non top-tier are 
present in bucket 3. In Table \ref{tab:correlation}, we provide the average values for these two features
for top-tier and non top-tier conferences within each bucket. We observe no clear correlation between publication count and feature
values. Note that these results are representative, i.e., this observation holds
true for all the features chosen.

\begin{table}[!thb]
\centering
 \caption{Average values for $\Delta$ CRDI(stddev) and $\Delta$
   EDI(stddev) for top-tier and non top-tier conferences in various
   buckets, created as per the average publication count of the conferences.}
 \begin{tabular}{|l|l|l|l|l|}
  \hline
Bucket (average & \multicolumn{2}{c|}{Avg. $\Delta$ CRDI(stddev)}& \multicolumn{2}{c|}{Avg. $\Delta$ EDI(stddev)} \\\cline{2-5}
publication count)& Top-tier   & Non top-tier  & Top-tier   & Non top-tier \\
\hline
Bucket1 (<35)    & 0.194    &0.189   &   0.522  & 0.334 \\
\hline
Bucket2 (35-150)    & 0.157  &0.146   &  0.226 & 0.319\\
\hline
Bucket3 (>150)    & 0.063  &0.228   &   0.334  & 0.346
\\
\hline
\end{tabular}
\label{tab:correlation}
\end{table}
\vspace{-0.2cm}

\subsection{Cross-correlation between features}
\begin{figure}[!thb]
  \centering
  \resizebox{.7\linewidth}{!}{\includegraphics{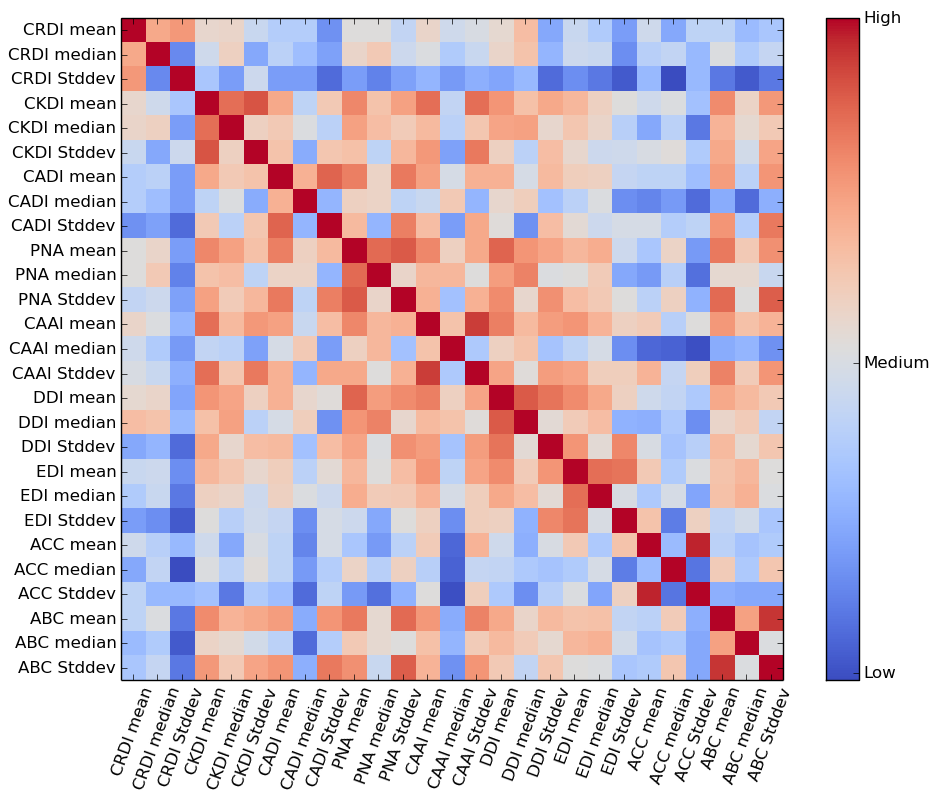}}
  \caption{Cross correlation between features: Red color represent highly correlated features (=1). Blue represent uncorrelated features (=0). Diagonal entries have maximum correlation (self) values = $1$. CKDI, CAAI, ACC and ABC have correlation values greater than 0.90 
between respective mean and standard deviation} \label{fig:correlation}
\end{figure}

Figure \ref{fig:correlation} presents cross-correlation between features. The correlation values lie between 0-1. Red color represent
highly correlated features (=1). Blue represent uncorrelated features (=0). Diagonal entries have maximum correlation (self) values = $1$.
We observe that mean and standard deviation of features correlates well. CKDI, CAAI, ACC and ABC have correlation values greater than 0.90 
between respective mean and standard deviation. Also, CAAI mean has maximum average correlation value of 0.67. It highly correlates with CKDI mean (0.85),
DDI mean (0.82) and PNA mean (0.80). Similarly, DDI mean highly correlates (0.87) with PNA mean. 

\vspace{-0.1cm}

\section{Experiments}
\label{sec:experiments}
In this section, we discuss in detail the experiments conducted to categorize the set of conferences into TT/NTT using
the feature set described above. Since \textit{ConfAssist} aims at
developing a conflict resolution framework, we use benchmark dataset, as
described in section \ref{sec:motivation} for these experiments. In this
benchmark, we had 53 conferences for which there was no conflict (set NC)
and 27 conferences for which conflict was observed (set CC). The main
idea behind this experiment was to study as to given a conference $x$
in the set CC, whether we can use the identified feature set to match
it with conferences in the set NC. Depending on whether the conference
$x$ matched more with the top-tier or non top-tier set, we 
predict a category for $x$.

We divide set NC randomly into training and validation subsets. Training set consists of 33 conferences (22 top-tier and 11 non top-tier). 
Validation set consists of 20 conferences (10 top-tier and 10 non top-tier). Support vector machine (SVM) with radial basis function (RBF) kernel is employed 
for conference categorization. We run grid search on validation set for optimal parameter ($\gamma$ and $C$) estimation with ten-fold cross validation. $\gamma$ denotes how far the influence of a single training example reaches, with low values indicating `far' and high values indicating `near'. More specifically, it represents inverse of the radius of influence of samples selected by the model as support vectors. Similarly $C$ denotes trade-offs between incorrect classification of training examples against simplicity of the decision surface. A low $C$ makes the decision surface smooth, while a high $C$ aims at classifying all training examples correctly by giving the model freedom to select more samples as support vectors.
Further, the estimated parameters, $\gamma=9.99^{-8}$ and $C=10^8$) are used for training the SVM model. 

\begin{figure}[!thb]
  \centering
  \resizebox{1\linewidth}{!}{\includegraphics{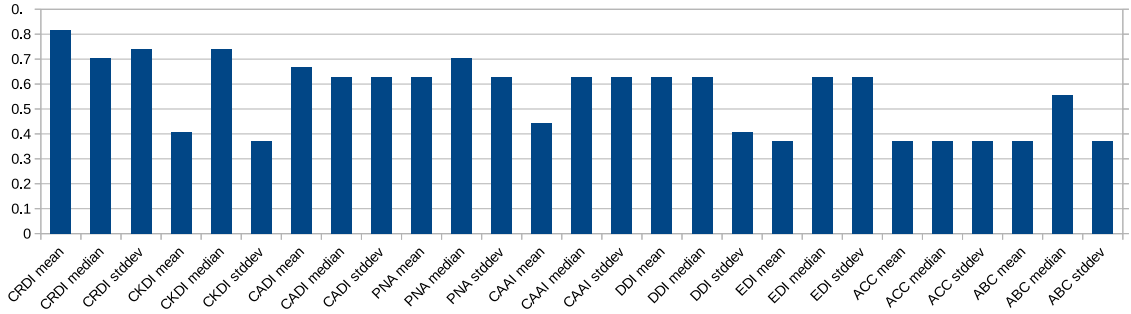}}
  \caption{Feature-wise SVM classification accuracy: CRDI mean performs the best with 81\% accuracy followed by CRDI standard deviation and CKDI median.} \label{featurewise_accuracy}
\end{figure}

\begin{figure}[!thb]
  \centering
  \resizebox{.9\linewidth}{!}{\includegraphics{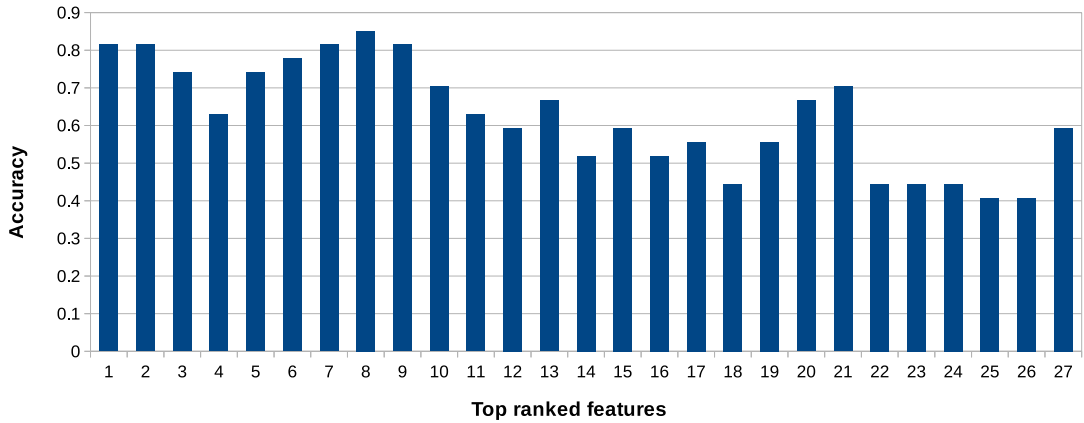}}
  \caption{Accuracy values by combining features together based on the accuracy rank list one at a time} \label{featurewise_ranks}
\end{figure}

We experiment with each feature individually to train the classification model. Figure \ref{featurewise_accuracy} presents accuracy for each feature. 
CRDI mean performs the best with 81\% accuracy followed by CRDI standard deviation and CKDI median. Co-authorship network features 
do not perform well as compared to the diversity based features. Further, we rank each feature based on the individual classification accuracy values. 
Figure \ref{featurewise_ranks} shows accuracy values by combining features together. We combine features based on the accuracy rank list (Table \ref{tab:ordered_list}) one at a time.
The system performs best with 85.18\% on eight features namely, CRDI mean, CRDI stddev, CKDI median, CRDI median, PNA median, CADI mean, CADI median and CADI stddev. However, as noted from Figure \ref{fig:correlation}, features CRDI mean and CRDI stddev are highly correlated, and so are CADI mean and CADI stddev. Therefore, in the final model, we exclude two redundant features, CRDI stddev and CADI stddev, and this provides an accuracy of 85.18\%, similar to that obtained while including these features.

\vspace{-0.1cm}

\begin{table}[!thb]
 \caption{ Ordered list of features as per prediction accuracy.}
 \resizebox{.9\textwidth}{!}{\begin{minipage}{\textwidth}
 \begin{tabular}{|l|l|l|l|l|l|l|l|l|l|l|l|}
\hline
1&CRDI mean&6&CADI mean&11&CAAI median&16&EDI stddev&21&CKDI stddev&26&ABC mean\\\hline
2&CRDI stddev&7&CADI median&12&CAAI stddev&17&ABC median&22&EDI mean&27&ABC stddev\\\hline
3&CKDI median&8&CADI stddev&13&DDI mean&18&CAAI mean&23&ACC mean&\multicolumn{2}{c}{}\\\cline{1-10}
4&CRDI median&9&PNA mean&14&DDI median&19&CKDI mean&24&ACC median&\multicolumn{2}{c}{}\\\cline{1-10}
5&PNA median&10&PNA stddev&15&EDI median&20&DDI stddev&25&ACC stddev&\multicolumn{2}{c}{}\\\cline{1-10}

  \end{tabular}
  \end{minipage}}
\label{tab:ordered_list}
\end{table}

Table \ref{survey} presents comparison between SVM results and the online survey. 
First column lists conference names. Second column presents SVM
 classification results. Third and fourth columns present total
 votes received and percentage of top-tier votes respectively. The last column shows if the majority votes for that conference
agree with the SVM results. Considering the majority voting for each conference, 23 out
of 27 (85.18\%) conferences were correctly classified, 3 were
incorrectly classified (ISSAC, DATE, and JCDL), 
while 1 got equal number of matching and non-matching votes (ICALP).

\begin{table*}[!thb]
 \caption{Comparing SVM results with the online survey: First column
   lists conference names. Second column presents SVM 
 classification results. Third and fourth columns present total
 votes received and percentage of top-tier votes. Last column
 shows agreement of survey results with the SVM results.}
 \resizebox{.83\textwidth}{!}{\begin{minipage}{\textwidth}
 \begin{tabular}{|l|c|c|c|c|c|}
  \hline
\textbf{Conference Name}&\parbox[t]{1.7cm}{\centering\textbf{SVM Class   (TT/NTT)}}&\parbox[t]{.7cm}{\centering\textbf{Total votes}}&\parbox[t]{1.5cm}{\centering\textbf{TT votes (\%)}}&\textbf{Agreement}\\
\hline
ACM Symposium on Parallel Algorithms and Architectures - SPAA&TT&13&62&Yes\\
\hline
International Symposium on Algorithms and Computation - ISAAC&TT&15&20&\textbf{No}\\
\hline
Design Automation Conference - DAC&TT&17&\textbf{94}&Yes\\
\hline
European Conference on Object-Oriented Programming - ECOOP&TT&14&21&Yes\\
\hline
Design, Automation, and Test in Europe - DATE&TT&17&47&\textbf{No}\\
\hline
International Colloquium on Automata, Languages and Programming - ICALP&TT&16&50&\textbf{Tie}\\
\hline
International Conference on Computer Aided Design - ICCAD&TT&12&92&Yes\\
\hline
International Conference on Distributed Computing Systems - ICDCS&TT&16&56&Yes\\
\hline
International Conference on Information and Knowledge Management - CIKM&TT&19&63&Yes\\
\hline
International Conference on Parallel Processing - ICPP&TT&12&75&Yes\\
\hline
Theory and Application of Cryptographic Techniques - EUROCRYPT&TT&20&65&Yes\\
\hline
International Conference on Robotics and Automation - ICRA&TT&15&87&Yes\\
\hline
Principles and Practice of Constraint Programming - CP&TT&11&64&Yes\\
\hline
ACM-IEEE Joint Conference on Digital Libraries - JCDL&NTT&21&62&\textbf{No}\\
\hline
Colloquium on Structural Information and Communication Complexity - SIROCCO&NTT&13&31&Yes\\
\hline
Foundations of Software Science and Computation Structure - FoSSaCS&NTT&15&33&Yes\\
\hline
Compiler Construction - CC&NTT&11&27&Yes\\
\hline
Data Compression Conference - DCC&NTT&14&43&Yes\\
\hline
European Symposium on Programming - ESOP&NTT&11&27&Yes\\
\hline
Fast Software Encryption - FSE&NTT&10&30&Yes\\
\hline
Field-Programmable Custom Computing Machines - FCCM&NTT&10&\textbf{0}&Yes\\
\hline
International Conference on Network Protocols - ICNP&NTT&13&46&Yes\\
\hline
Mathematical Foundations of Computer Science - MFCS&NTT&10&40&Yes\\
\hline
Network and Operating System Support for Digital Audio and Video - NOSSDAV&NTT&15&27&Yes\\
\hline
Symposium on Graph Drawing - GD&NTT&12&25&Yes\\
\hline
Applications of Natural Language to Data Bases - NLDB&NTT&15&40&Yes\\
\hline
Symposium on Theoretical Aspects of Computer Science - STACS&NTT&13&38&Yes\\
  \hline
  \end{tabular}
\end{minipage}}
\label{survey}
\end{table*}
\vspace{-0.2cm}
\section{Factor Analysis}
\label{sec:factor}
We further perform factor analysis to determine how many different groups of features are present in the data. We perform principal component analysis using \textit{Weka}\footnote{http://www.cs.waikato.ac.nz/ml/weka/} tool. The analysis resulted in 11 orthogonal factors. Table \ref{tab:PCA} presents factors in decreasing order of the ranks. We also create model after including top ranked features from each factor. However, the prediction accuracy drops from 85.18\% to 55.5\% (considering all 11 top ranked features) and to 62.9\% (best, considering only four top ranked features). 

\begin{table*}[!thb]

 \caption{Independent factors and corresponding ranks from PCA. Each row represents a factor alongwith top five features (separated from coefficient value by colon) in that factor.}
 \resizebox{.9\textwidth}{!}{\begin{minipage}{\textwidth}
 \begin{tabular}{|c|l|l|l|l|l|}
 \hline
Rank&\multicolumn{5}{c|}{Independent Factors}\\\hline
1&0.256:CAAI mean&+0.243:PNA mean&+0.241:DDI mean&+0.241:EDI mean&+0.24 :CAAI Stddev\\\hline
2&0.431:ABC mean&+0.405:ABC Stddev&+0.373:ACC mean&+0.322:ACC Stddev&+0.306:ABC median\\\hline
3&-0.345:EDI Stddev&+0.314:DDI median&+0.302:CAAI median&-0.291:CKDI Stddev&-0.282:ACC Stddev\\\hline
4&0.484:ACC Stddev&+0.438:ACC mean&-0.38:DDI Stddev&+0.286:CRDI Stddev&+0.27 :CRDI mean\\\hline
5&-0.392:ABC mean&-0.384:ABC Stddev&-0.278:CRDI median&+0.27 :CADI median&-0.27:CRDI mean\\\hline
6&-0.489:CADI Stddev&-0.376:CADI mean&+0.337:DDI median&+0.295:ABC median&-0.264:ABC Stddev\\\hline
7&-0.73:ACC median&-0.27:ABC median&+0.229:CADI median&+0.22 :CAAI median&-0.201:CKDI Stddev\\\hline
8&0.629:ABC median&+0.283:CKDI median&+0.266:CADI Stddev&-0.265:DDI median&-0.212:DDI Stddev\\\hline
9&0.425:CAAI median&-0.347:PNA Stddev&+0.312:ACC median&-0.311:CADI Stddev&+0.283:CADI median\\\hline
10&0.449:EDI median&-0.415:CAAI median&+0.358:CRDI median&+0.311:EDI mean&-0.276:CRDI Stddev\\\hline
11&0.553:CRDI Stddev&-0.402:CRDI median&+0.335:EDI median&+0.328:CADI median&-0.325:PNA median\\\hline
  \end{tabular}
\end{minipage}}
\label{tab:PCA}
\end{table*}
\vspace{-0.2cm}

\section{Conclusions and Future work}
\label{sec:conclusion}
In this paper, we aimed at identifying the underlying features that separate a top-tier conference from a non top-tier conference.
We started our study with a motivating experiment that shows while there are many 
clear cases where expert agreement can be almost immediately achieved as to whether 
a conference is a top-tier or not, there are equally many cases that can result in a conflict 
even among the experts. We presented a hypothesis that the top-tier
conferences are more stable than the non top-tier ones in maintaining
the similar level of diversity over the years. Accordingly, we identified nine distinct quantities, that gave us 27
different features (mean, median and standard deviation for each of
the quantities). From an analysis of the fluctuation patterns,
top-tier conferences were indeed found to be much more stable than
the non top-tier ones. We also presented comparison of dynamics of a
new conference with matured top-tier and non top-tier conferences,
which confirms that the proposed features can help in obtaining some initial signals
of future popularity of the new conference. We then used these features for the development of
\textit{ConfAssist}, which uses a SVM based classification framework
to classify the conflicting conferences into top-tier or non
top-tier categories. An online survey with 28 human experts produced
an agreement of 85.18\% (majority voting) over a set of 27 conferences. 
 
This study can be further extended to analyze other such features as
well as other venue types. Further, a similar study can be conducted for fields other than computer science
to understand whether this hypothesis holds true in general for any
venue. Moreover, if these features are known, they can add to our
understanding of the dynamics of research venues, as to what is the
underlying process behind a conference becoming a top-tier. On an
ambitious note, it might also prove beneficial to the older venues on
the verge of extinction and the newer venues coming into limelight.

\label{}

\bibliographystyle{elsarticle-harv}\biboptions{authoryear}
\bibliography{JOI.bib}

\end{document}